\documentclass[twocolumn]{aastex631}
\usepackage{hyperref}
\usepackage{floatrow}
\usepackage{amsmath}

\usepackage{tikz,xcolor}

\shorttitle{Double-peaked SN 2006aj}
\shortauthors{Zhang, Yu \& Liu}

\graphicspath{{./}{figures/}}

\begin{document}

\title{The effects of a magnetar engine on the gamma-ray burst-associated supernovae: Application to double-peaked SN 2006aj}

\author{Zhen-Dong Zhang}
\affiliation{Institute of Astrophysics, Central China Normal University, Wuhan 430079, China; \href{mailto:yuyw@ccnu.edu.cn}{yuyw@ccnu.edu.cn}}

\author{Yun-Wei Yu}
\affiliation{Institute of Astrophysics, Central China Normal University, Wuhan 430079, China; \href{mailto:yuyw@ccnu.edu.cn}{yuyw@ccnu.edu.cn}}
\affiliation{Key Laboratory of Quark and Lepton Physics (Central China Normal University), Ministry of Education, Wuhan 430079, China}

\author{Liang-Duan Liu}
\affiliation{Institute of Astrophysics, Central China Normal University, Wuhan 430079, China; \href{mailto:yuyw@ccnu.edu.cn}{yuyw@ccnu.edu.cn}}
\affiliation{Key Laboratory of Quark and Lepton Physics (Central China Normal University), Ministry of Education, Wuhan 430079, China}

\begin{abstract}
A millisecond magnetar engine has been widely suggested to exist in gamma-ray burst (GRB) phenomena, in view of its substantial influences on the GRB afterglow emission. In this paper, we investigate the effects of the magnetar engine on the supernova (SN) emission which is associated with long GRBs and, specifically, confront the model with the observational data of SN 2006aj/GRB 060218. SN 2006aj is featured by its remarkable double-peaked ultraviolet-optical (UV-opt) light curves. By fitting these light curves, we demonstrate that the first peak can be well accounted for by the breakout emission of the shock driven by the magnetar wind, while the primary supernova emission is also partly powered by the energy injection from the magnetar. The magnetic field strength of the magnetar is constrained to be $\sim 10^{15}$ G, which is in good agreement with the common results inferred from the afterglow emission of long GRBs. In more detail, it is further suggested that the UV excess in the late emission of the supernova could also be due to the leakage of the non-thermal emission of the pulsar wind nebula (PWN), if some special conditions can be satisfied. The consistency between the model and the SN 2006aj observation indicates that the magnetar engine is likely to be ubiquitous in the GRB phenomena and even further intensify their connection with the phenomena of superluminous supernovae.

\end{abstract}


\section{Introduction} \label{sec:intro}

One of the unresolved mysteries of gamma-ray burst (GRB) phenomena is the nature of their central engines, which could be a black hole (BH) or a neutron star \citep[NS;][]{1993ApJ...405..273W,1992Natur.357..472U,1992ApJ...392L...9D}. Since the discovery of the shallow-decay and plateau afterglows and, in particular, afterglow flares, it has been widely suggested that a rapidly rotating and highly magnetized remnant NS (i.e., millisecond magnetar) should play a crucial role in causing these afterglow features \citep{1998A&A...333L..87D,1998MNRAS.298...87D,2001ApJ...552L..35Z,2006MNRAS.372L..19F,2006Sci...311.1127D,2007A&A...470..119Y,2007MNRAS.377.1638D,2007ApJ...659..561M,2007ApJ...665..599T,2010MNRAS.402..705L,2010ApJ...715..477Y}.
It is worth to mention that millisecond magnetars could also be able to drive an unusual type of supernovae called super-luminous supernovae (SLSNe; at least, a part of them),
which are about $10-100$ times brighter than normal SNe \citep{2010ApJ...717..245K,2010ApJ...719L.204W,2012Sci...337..927G,2013ApJ...770..128I,2017ApJ...840...12Y,2017ApJ...842...26L}. Then, in view of that the long GRBs are also originated from the collapse of massive stars, it is of significance to ask what imprints the post-GRB magnetars can leave in their supernova counterparts. In other words, what we concern in this paper is, besides the GRB afterglow features, what else signatures of the post-GRB magnetars exist in the GRB-associated supernovae (GRBSNe).

Generally speaking, the magnetic field strengths of the magnetars in long GRBs are usually found to be about an order of magnitude higher than those in the SLSN cases \citep{2017ApJ...840...12Y}. Then, the spin-down timescale of these magnetars is usually much shorter than the diffusion timescale of the SN ejecta. Therefore, different from SLSNe, it is expected that GRBSNe cannot be brightened to be much more luminous than normal core-collapse supernovae and, instead, the majority of the spin-down energy is primarily converted into the kinetic energy of the supernova ejecta.
 In observations, since the discovery of the GRB 980425/SN 1998bw association event \citep{1998Natur.395..670G,1998Natur.395..672I}, several tens of GRBSNe have been discovered (see, \citealt{2012grb..book..169H,2017AdAst2017E...5C,2018ApJ...862..130L} and the references therein).
On the one hand, the spectral features of these GRBSNe indicate they belong to broad-lined Type Ic supernovae \citep[SNe Ic-BL;][]{2006ARA&A..44..507W,2017AdAst2017E...5C}, which hints the GRBSNe are indeed given huge kinetic energy. In the light of statistics, it is found that the kinetic energy of GRBSNe is clustered around $E_{\rm k} \sim (1-2) \times 10^{52} \  \rm{erg}$ \citep{2014MNRAS.443...67M}, which is in good agreement with the typical rotational energy of a millisecond magnetar. On the other hand, although the GRBSNe are not expected to be as luminous as SLSNe\footnote{SN 2011kl is an exception, which was associated with ultra-long GRB 111209A and simultaneously can be classified as a SLSN \citep{2013ApJ...766...30G,2013ApJ...779...66S}. This somehow indicates that SLSNe and long GRBs could have a unified origin and there is a continuous transition between the two phenomena \citep{2017ApJ...840...12Y}. GRB 111209A/SN 2011kl is just a sample on the critical line.}, their luminosities are still found to be somewhat higher than those of normal supernovae and thus a more powerful energy source is usually required. The most direct explanation of such an energy source is a relatively high mass of $\rm ^{56}Ni$ \citep{2011ApJ...741...97D,2016MNRAS.457..328L,2016MNRAS.458.2973P,2018ApJ...862..130L}, which could in principle be synthesized during the shock acceleration of the SN ejecta due to the energy injection from the magnetar \citep{2015MNRAS.451..282S,2015ApJ...810..109N}. In summary, the basic features of GRBSNe can generally be consistent with the millisecond magnetar engine model.

To be specific, the energy injection from a remnant magnetar to SN ejecta is through a relativistic leptonic wind, which can lastingly push the ejecta and drive a pair of shocks. While the reverse shock terminates the injecting wind material persistently, the forward shock would quickly cross the ejecta. Then, the emission due to these shocks could provide potential observational signatures of the existence of the remnant magnetar. First of all,  the breakout of the forward shock emission could produce an early peak in the supernova light curve before its primary peak, which has indeed been found in several SLSNe recently \citep{2016ApJ...821...36K,2021ApJ...911..142L}. So, considering that the spin-down luminosity of GRB magnetars can be much higher than those of the SLSNe, it is natural to expect that a more significant shock breakout (SBO) emission peak can also appear in the GRBSN light curves. This expectation leads us to recall the famous GRBSN: SN 2006aj, which was associated with GRB 060218 and impressive with its double-peaked light curves \citep{2006Natur.442.1008C,2006Natur.442.1018M,2006Natur.442.1011P,2006Natur.442.1014S,2007ApJ...654..385K}. 
The primary purpose of this paper is to test whether the early UV-optical peak of SN 2006aj can be contributed by the SBO due to the injection of a magnetar wind.

The paper is organized as follows. The basic properties of SN 2006aj are introduced in the next section. In Section \ref{sec:model}, we reproduce the double-peaked light curves of SN 2006aj by using the model including a wind-driven SBO emission and a magnetar-aided supernova emission. In Section \ref{sec:PWNmodel}, for a more elaborate fitting, we further consider the possible leakage of the non-thermal emission from the shocked wind region which is termed as usual as a pulsar wind nebula (PWN). Finally, we give a summary and conclusions in Section \ref{sec:summary}.

\begin{figure*}[htbp]
	\centering
	\includegraphics[width=0.7\textwidth]{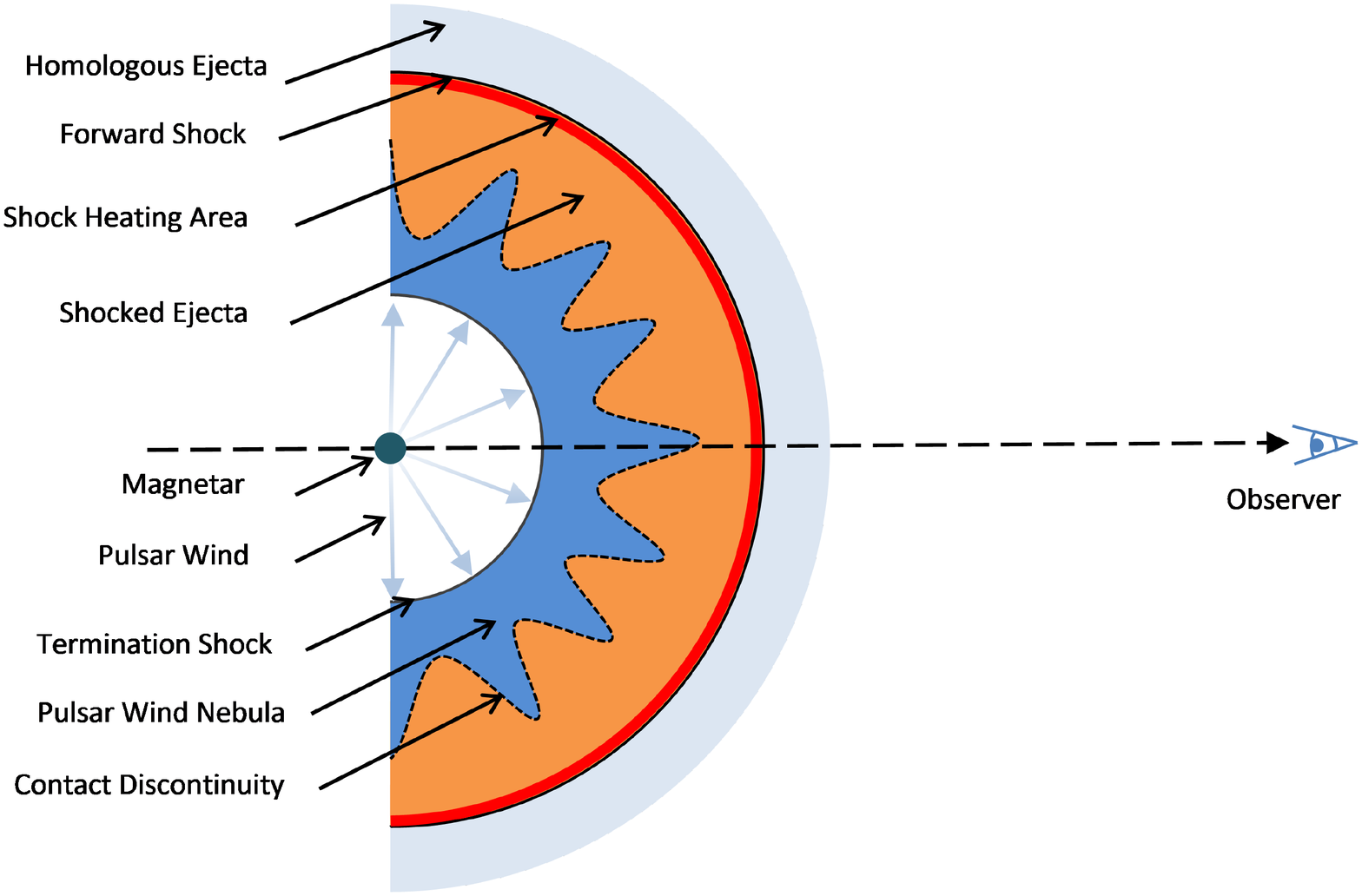} \\
	\caption{The schematic diagram of the model (not to scale). }
	\label{ske1}
\end{figure*}

\section{GRB 060218/SN 2006aj} \label{sec:sample}
GRB 060218 was detected on 2006 February 18 by \emph{Swift} satellite \citep{2004ApJ...611.1005G}, which is famous by its ultra-long and X-ray prompt emission including a blackbody component. Thanks to the long duration and the near distance of a redshift $z \approx 0.0335$ \citep{2006GCN..4786....1C,2006ApJ...643L..99M}, its multi-wavelength afterglows and as well as the associated supernova SN 2006aj had been monitored carefully and deeply. One of the most mysterious features of the afterglow emission is the appearance of a big bump in the UV-optical light curves before the normal supernova peak \citep{2006Natur.442.1008C,2006A&A...457..857F,2006ApJ...645L..21M,2006A&A...454..503S,2010A&A...523A..56S,2019MNRAS.484.5484E}. As the first impression, this UV-optical peak is considered to be the tail of the blackbody component of the prompt X-ray emission \citep{2006Natur.442.1008C}. However, \cite{2007MNRAS.375L..36G} found that it is difficult to connect the UV-optical peak with the thermal X-rays by a naturally evolving blackbody emission. Alternatively, they explained the overall spectrum from the optical-UV to the X-ray by a synchrotron self-Compton model, which is however disfavored by the radio observations \citep{2006Natur.442.1014S,2016MNRAS.460.1680I}.

In principle, the breakout emission of a supernova bounce shock could be a promising origin for the early UV-optical peak. However, the early peak in the GRB 060218/SN 2006aj event is obviously too bright to be explained by a normal supernova SBO. Then, alternatively, it was suggested that the UV-optical peak could be contributed by a shock-heated wind \citep{2007ApJ...667..351W,2016MNRAS.460.1680I} or a SBO driven by a jet that was choked in a low-mass and extended envelope around the SN ejecta \citep{2015ApJ...807..172N}. However, both of these scenarios require an elaborate progenitor structure. For GRB 060218/SN 2006aj, the mass of its progenitor star is constrained to be around $\sim 20  \ M_{\odot}$ \citep{2006Natur.442.1018M}, which is likely to give birth to a NS remnant rather than a BH. Furthermore, a NS remnant is also required to account for the production of a relatively high mass ($\sim 0.05 \ M_{\odot}$) of neutron-rich $\rm ^{58}Ni$ \citep{2007ApJ...658L...5M}, which is inferred from the corresponding emission line. Therefore, it is reasonable and necessary to investigate whether the SBO emission driven by a magnetar wind can provide an explanation for the UV-optical peak of SN 2006aj and, moreover, what constraints can be obtained on the magnetar parameters.

\section{The magnetar wind-driven SBO } \label{sec:model}
\subsection{Basic equations}
Homologous expanding supernova ejecta is considered to surround a remnant NS. The spin-down of the NS drives a relativistic wind to collide with the supernova ejecta. As a result, a forward shock is formed to propagate into the supernova ejecta, while the wind material is terminated by the reverse shock. The SBO emission is produced when the forward shock is very close to the outer surface of the supernova ejecta. Our model is depicted in Figure \ref{ske1}.

Here we adopt the model developed in \cite{2016ApJ...819..120L} to describe the dynamical evolution and emission of the SBO and supernova. According to the energy conservation, the dynamical evolution of the forward shock is determined by
\begin{eqnarray}
\frac{dv_{\rm sh}}{dt}&=&{1\over{M_{\rm sh}v_{\rm sh}}}\left[(L_{\rm
inj}-L_{\rm sbo}-L_{\rm sn})\right.\nonumber \\
&&\left.-{1\over2}\left(v_{\rm sh}^2-v_{\rm
ej}^2\right)\frac{dM_{\rm sh}}{dt}- \frac{dU_{\rm}^{}}{dt^{}}\right],\\
\frac{dM_{\rm sh}}{dt}&=&4\pi R_{\rm sh}^{2}\left(v_{\rm sh}-v_{\rm
ej}\right)\rho_{\rm ej},
\end{eqnarray}
where $v_{\rm sh}$ and $R_{\rm sh}$ are the velocity and radius of the forward shock, $M_{\rm sh}$ is the total mass of the shocked ejecta, $L_{\rm inj}$ is the power of the energy sources, $L_{\rm sbo}$ and $L_{\rm sn}$ are the luminosities of the SBO emission and supernova emission, $v_{\rm ej}$ and $\rho_{\rm ej}$ are the velocity and density of the ejecta material just in front of the shock. $U$ is the total internal energy of the supernova ejecta, which can be divided into the shock-accumulated part ($U_{\rm sh}$) and the remaining part ($\tilde{U}=U-U_{\rm sh}$). The evolution of these internal energies can be coupled with the dynamical quantities by the following two equations:
\begin{eqnarray}
\frac{dU_{\rm sh}^{}}{dt^{}}&=&H_{\rm sh}^{}-{P_{\rm sh}^{}
}\frac{d (\epsilon V_{\rm}^{})}{dt^{}}-L_{\rm sbo}^{},
\\
\frac{d\tilde{U}^{}}{dt^{}}&=&L_{\rm
inj}^{}-\tilde{P}^{}\frac{dV_{\rm}^{}}{dt^{}}-L_{\rm sn}^{},
\end{eqnarray}
where the shock heating rate is given by $H_{\rm sh}^{}={1\over 2}\left[v_{\rm sh}-v_{\rm ej}(r_{\rm
sh},t)\right]^2(dM_{\rm sh}/dt)$, the pressures read $P_{\rm sh}^{} =U_{\rm sh}^{}/(3\epsilon V_{\rm}^{})$ and $\tilde{P}^{}=\tilde{U}^{}/3V_{\rm}^{}$, and $V$ is the volume of the shocked ejecta. The parameter $\epsilon$ represents the volume fraction of the shock heating region in the entire shocked ejecta, which is considered to be extremely small.

Finally, the emission luminosity of the SBO and the supernova can be given by \citep{2016ApJ...819..120L}:
\begin{equation}
	L_{\rm sbo}=\frac{R_{\rm max}^2 U_{\rm sh} c}{\epsilon R_{\rm sh}^3+R_{\rm max}^3-R_{\rm sh}^3} \left[\frac {1-e^{-(\epsilon \tau_{\rm sh}+\tau_{\rm un})}} {\epsilon \tau_{\rm sh}+\tau_{\rm un}}\right],\label{Lsbo}
\end{equation}
and
\begin{equation}
	L_{\rm sn}=\frac{\tilde{U} c}{R_{\rm max}} \left[\frac {1-e^{-( \tau_{\rm sh}+\tau_{\rm un})}} { \tau_{\rm sh}+\tau_{\rm un}}\right],\label{Lsn}
\end{equation}
respectively, where $\tau_{\rm sh}= {3\kappa M_{\rm ej}}/{4\pi R_{\rm sh}^{2} }$ and
$\tau_{\rm un}=  \int_{R_{\rm sh}}^{R_{\max}} \kappa \rho_{\rm ej}
dr$ are the optical depths of the shocked and unshocked ejecta, and $\kappa=0.1\rm cm^{2}g^{-1}$ is the opacity. In order to calculate the multi-band light curves, we simply assume a blackbody spectrum for the SBO and supernova emission, which has
a temperature of 
\begin{equation}
	T_{\rm{eff}}=\left (\frac{L_{\rm e}}{4 \pi \sigma_{\rm SB} R_{\rm ph}^{2}}\right)^{1 / 4},\label{Teff}
\end{equation}
where $L_{\rm e}=L_{\rm sbo}+L_{\rm sn}$, $\sigma_{\rm SB}$ is the Stephan-Boltzmann constant, and $R_{\rm ph}$ is the photosphere radius where the electron-scattering optical depth satisfies
\begin{equation}
\tau_{\rm ph}=\int_{R_{\rm ph}}^{R_{\max}} \kappa \rho_{\rm ej}dr={2\over 3}.
\end{equation}
If $R_{\rm ph} < R_{\rm sh}$, we simply set $R_{\rm ph} = R_{\rm sh}$.

\subsection{Physical inputs} \label{subsec:ejecta}
Here we consider that GRBSNe could be powered by the radioactivity of $^{56}$Ni and the spin-down of a remnant NS.
The heating power due to the radioactive decays can be written as \citep{1982ApJ...253..785A,2008MNRAS.383.1485V}:
\begin{equation}
	L_{\rm r}(t)= M_{\rm Ni} \left[(\epsilon_{\rm Ni}-\epsilon_{\rm Co}) e^{-t/\tau_{\rm Ni}}+ \epsilon_{\rm Co} e^ {-t/\tau_{\rm Co}}\right],
\end{equation}
where $M_{\rm Ni}$ is the total mass of $^{56}$Ni, $\epsilon_{\rm Ni} = 3.90 \times 10^{10} \ \rm{erg \ s^{-1} \ g^{-1}}$, $\epsilon_{\rm Co} = 6.78 \times 10^{9}  \ \rm{erg \ s^{-1} \ g^{-1}} $, 
$\tau_{\rm Ni}= 8.77 \ \rm{days}$, and $\tau_{\rm Co}= 111.3 \ \rm{days}$. 
Meanwhile, the spin-down power of the remnant NS by the dipole radiation is written as:
\begin{equation}
L_{\rm sd}(t)= L_{\rm sd,0} \left(1+\frac{t}{t_{\rm sd}} \right)^{-2},
\label{eq:sd}
\end{equation}
where $L_{\rm sd,0}$ is the initial value of the luminosity and $t_{\rm sd}$ represents the spin-down timescale. In our calculations, we take $L_{\rm inj}=L_{\rm r}+L_{\rm sd}$.

For the SN ejecta, we take a density distribution as \citep{2018ApJ...868L..24L}:
\begin{equation}
	\rho_{\rm ej}(x,t)= \rho_{\rm 0} \eta(x)  \left[\frac{R_{\rm max,0}-R_{\rm min,0}}{R_{\rm max}(t)-R_{\rm min}(t)} \right]^3,
	\label{eq:rho}
\end{equation}
where $R_{\rm min}(t) = R_{\rm min,0} + v_{\rm min} t$ and $R_{\rm max}(t) = R_{\rm max,0} + v_{\rm max} t$ are the evolving minimum and maximum radius of the ejecta with $v_{\rm min}$ and $v_{\max}$ being the corresponding velocities and the subscript $\mathrm{0}$ represents their initial values. The density profile is described by a broken power-law as
\begin{equation}
 \eta(x)=\left\{
\begin{array}{rcl}
	\left(\frac{x}{x_0} \right)^{-n},    &      & {x_0 \leqslant x \leqslant 1},\\
	\left(\frac{x}{x_0} \right)^{-\sigma},    &      & {0 \leqslant x < x_0},\\
\end{array} \right.
\end{equation}
with
\begin{equation}
	 x=\frac{R-R_{\rm min}}{R_{\rm max}-R_{\rm min}},
	 \label{eq:x}
\end{equation}
where the power-law indices satisfy $n>5$ and $\sigma<3 $.
Denoting the mass and kinetic energy of the ejecta by $M_{\rm ej}$ and $E_{\rm k}$, we can obtain \citep{2004A&A...427..453V,2018ApJ...868L..24L}
\begin{equation}
	\rho_{\rm 0} ={ M_{\rm ej}\over 4 \pi I_{\rm m} R_{\rm max,0}^3},
\end{equation}
and
\begin{equation}
v_{\max} = \sqrt{\frac{2 E_{\rm k} I_{\rm m}}{M_{\rm ej} I_{\rm k}}},
\end{equation}
where
\begin{eqnarray}
	 I_{\rm m} &=& \frac{1}{3-\sigma} x_0^3+ \frac{1}{3-n}(x_0^n-x_0^3),\\
	I_{\rm k} &=& \frac{1}{5-\sigma} x_0^5+ \frac{1}{5-n}(x_0^n-x_0^5),
\end{eqnarray}
and $x_0=0.1$ is taken in our calculations. The values of $R_{\rm min,0}$ and $v_{\rm min}$, as they are sufficiently small, will not significantly influence the following calculation results.

\begin{figure*}[htbp]
	\centering
	\includegraphics[width=1\textwidth]{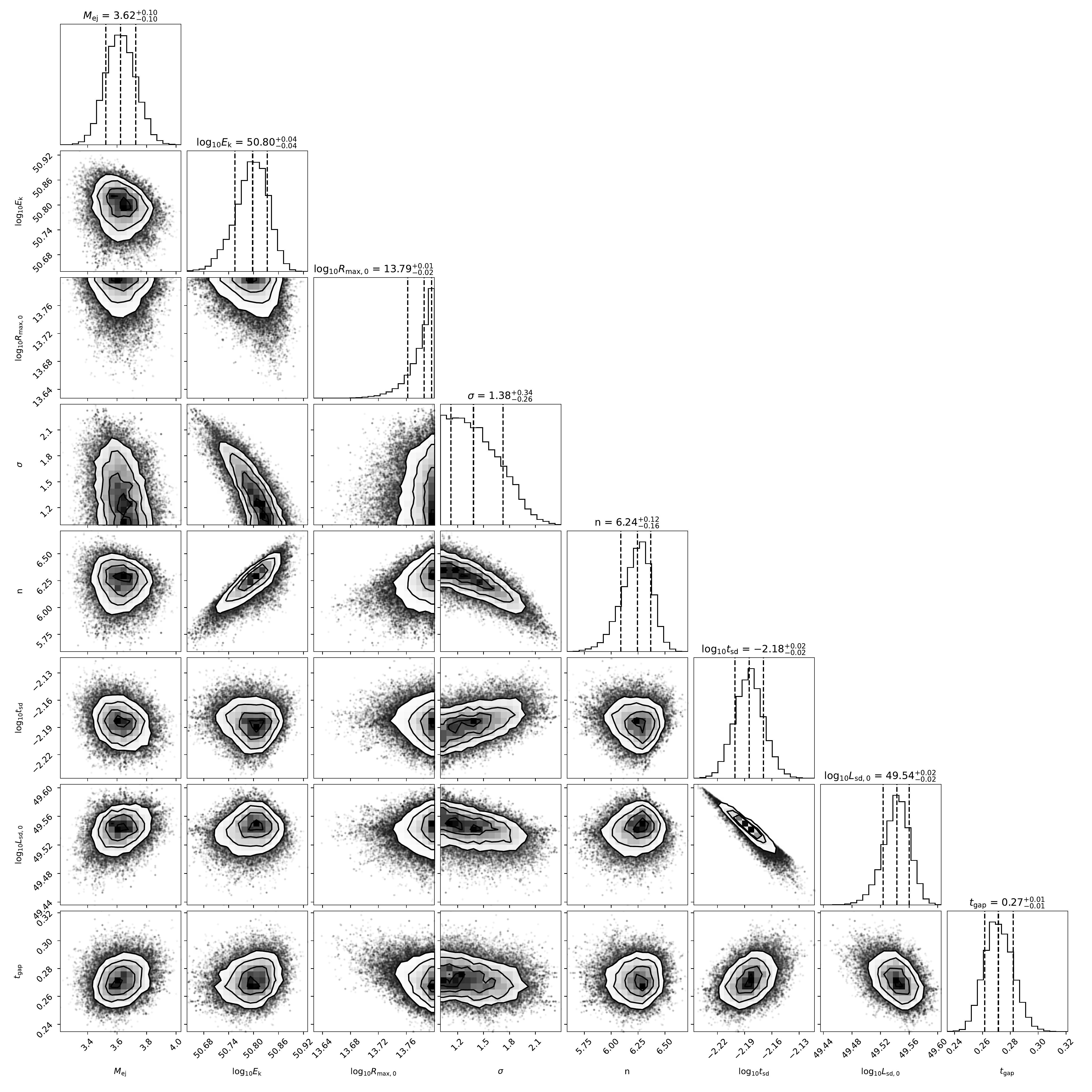} \\
	\caption{Posterior distributions of parameters for the fittings of SN 2006aj.}
	\label{MCMC1}
\end{figure*}

\subsection{Results and discussion} \label{sec:results}

\begin{table}[t]
	\centering
	\setlength{\tabcolsep}{2mm}{}
	\renewcommand\arraystretch{1.4}
	\begin{tabular}{lllll}
		\hline \hline Parameter & Value & Best-fit & Range \\
		\hline
		$M_{\rm ej} \ (M_{\odot})$ &  $3.62_{-0.10}^{+0.10} $ & 3.64  & (1,10)\\
		$\log_{10}E_{\rm k} \ (\rm{erg})$ & $50.80_{-0.04}^{+0.04}$ & 50.82 & (50,52) \\
		$\log_{10}R_{\rm{max,0}} \  (\rm{cm})$&   $13.79_{-0.02}^{+0.01}$ & 13.80 & (11,13.8) \\
		$\sigma$ & $1.38_{-0.26}^{+0.34}$ & 1.09 & (1,3)\\
		$n$ & $6.24_{-0.16}^{+0.12}$ & 6.38 & (5,10)\\
		$\log_{10}t_{\rm sd} \  (\rm{day})$ & $-2.18_{-0.02}^{+0.02}$ & -2.21 & (-4,0) \\
		$\log_{10}L_{\rm sd,0} \  (\rm{erg/s})$ & $49.54_{-0.02}^{+0.02}$ & 49.57 & (47,50)\\
		$t_{\rm gap} \ (\rm{day})$ &  $0.27_{-0.01}^{+0.01}$ & 0.27 & (0,2)\\

		$M_{\rm Ni} \ (M_{\odot})$ & 0.1 (fixed) \\
		
		\hline
		$E_{\rm sd} \ (\rm{erg})$ &  & $1.988 \times 10^{52}$ & \\
		$P_{0} \  (\rm{ms})$ &  & 1.003 & \\
		$B_{\rm p} \ (10^{14}\rm{G})$ &  & 9.694 & \\
		\hline
		
	\end{tabular}\\
	
	\caption{Parameters estimated from the modeling of SN 2006aj by MCMC-sampling.} \label{tab1}
\end{table}

\begin{figure*}[tp]
	\centering
	\includegraphics[width=0.49\textwidth]{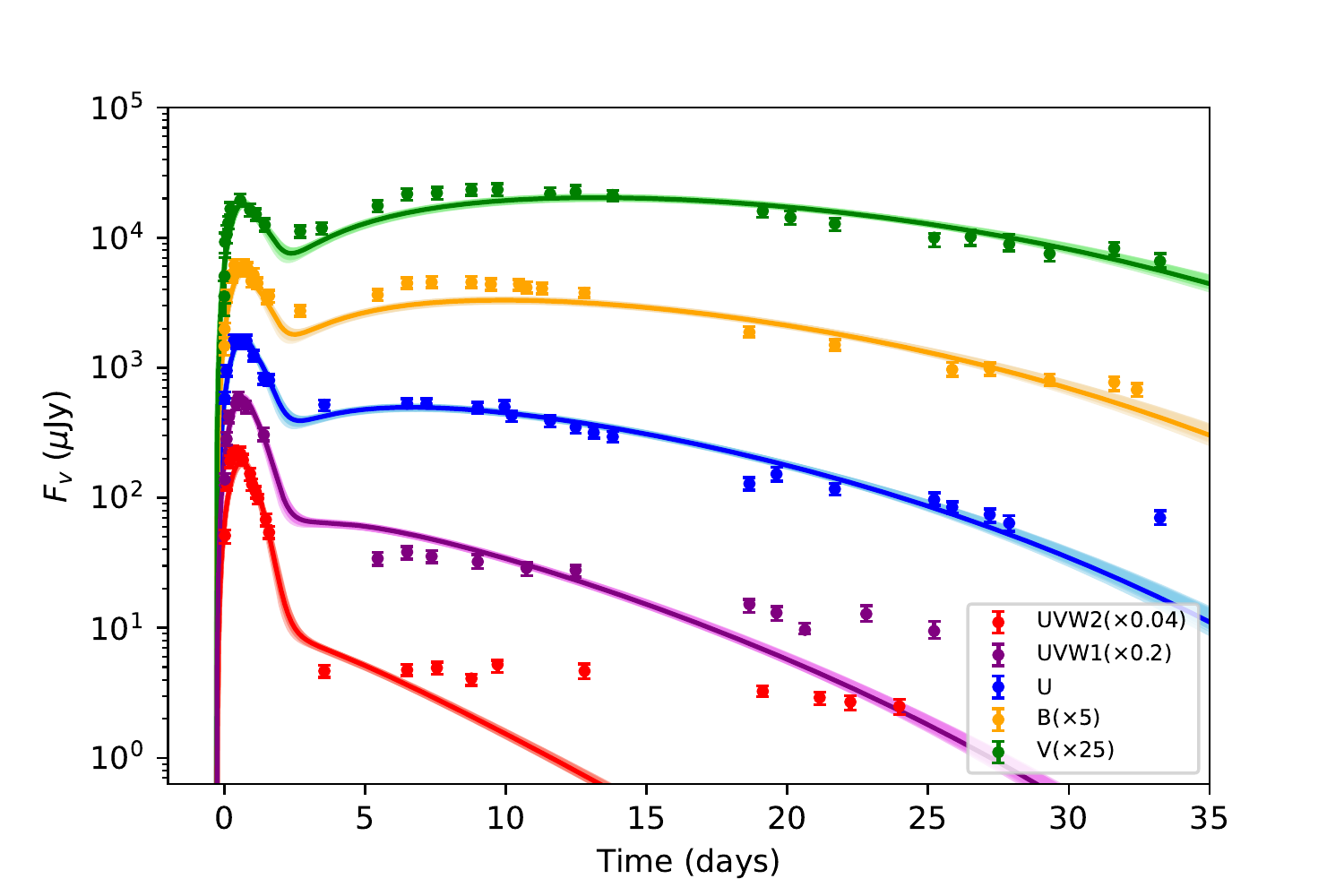} \includegraphics[width=0.49\textwidth]{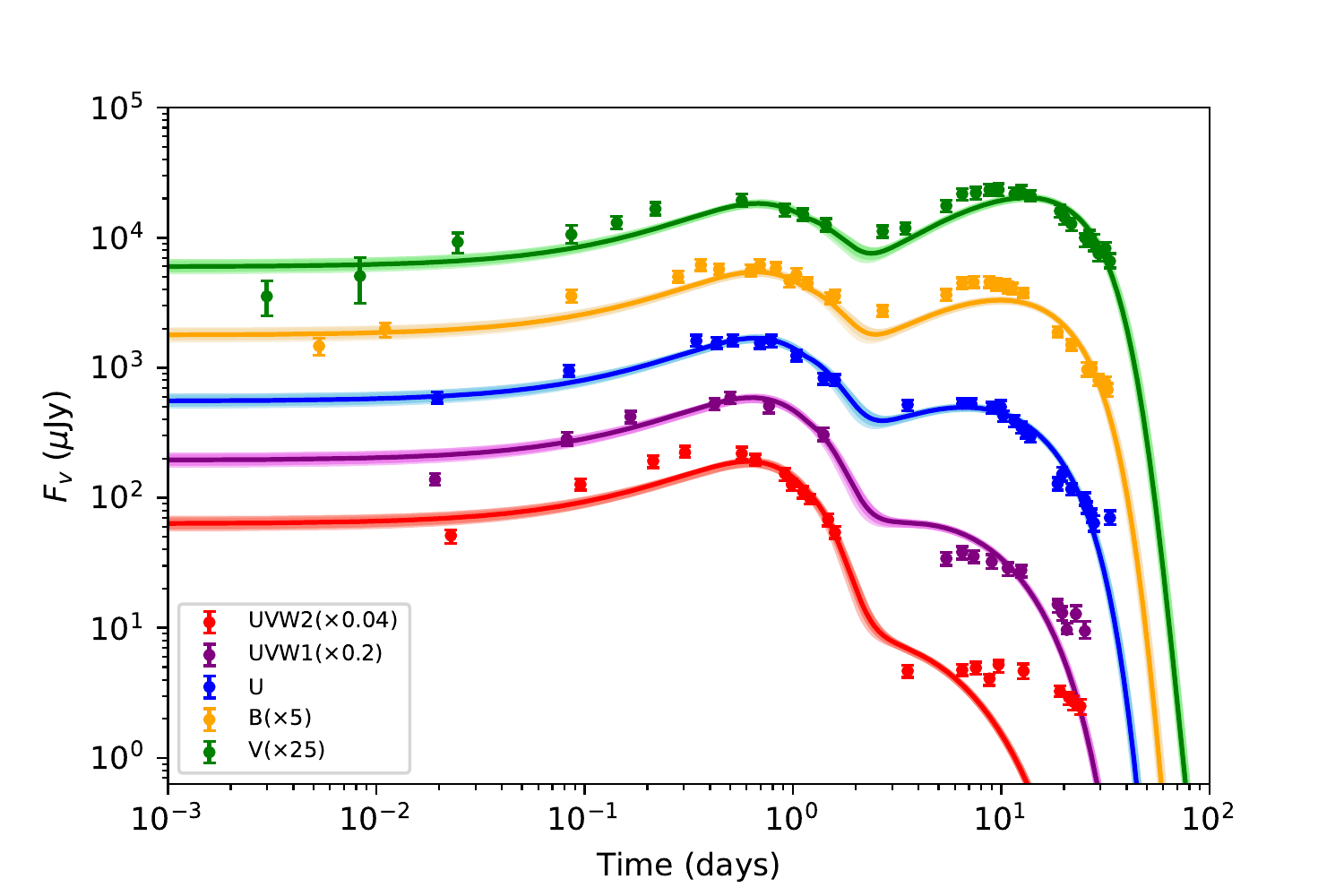}\\
	\caption{The best fit of the multi-wavelength double-peaked light curves of SN 2006aj by using the model including a magnetar wind-driven SBO emission and a magnetar-aided supernova emission, for the parameter values listed in Table \ref{tab1}. The left and right panels are shown in a linear and logarithmic-scale x-axis, respectively. The observational data are taken from \cite{2015ApJ...807..172N} and \cite{2006Natur.442.1008C}.}
	\label{LC1}
\end{figure*}

\begin{figure}[htbp]
	\centering
	\includegraphics[width=\textwidth]{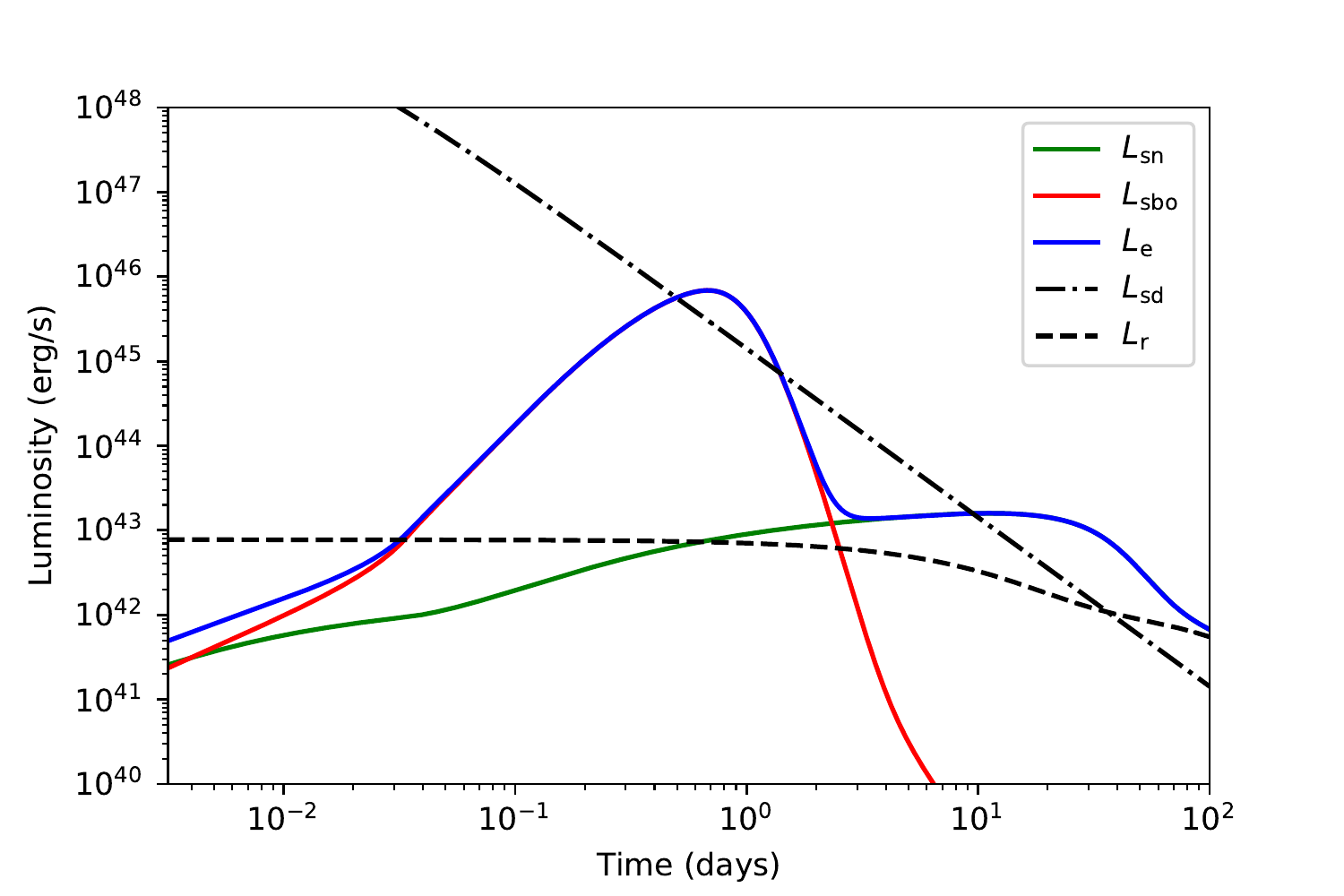}
	\caption{The bolometric light curves of the SBO and supernova emission in comparison with the luminosity evolution of the heating powers. The parameter values are same to  Figure \ref{LC1}. The zero-time in this figure is $t_{\rm mag,0}$.}
	\label{bol1}
\end{figure}

By substituting the heating powers and the ejecta profile into the model equations, we obtain the temporal evolution of $L_{\rm sbo}$ and $L_{\rm sn}$ and then fit the multi-band light curves of SN 2006aj. Then, the model parameters are constrained by MCMC sampling with $\emph{emcee}$ package \citep{2013PASP..125..306F}. The parameter constraints with
$1\sigma$ confidence level are presented in Figure \ref{MCMC1} and Table \ref{tab1}. The corresponding fitting results are shown in Figure \ref{LC1} and, furthermore, we plot the bolometric light curve for the best-fit parameters in Figure \ref{bol1}. As shown, the peak bolometric luminosity of the SBO emission $L_{\rm sbo}$ could reach $\sim 10^{46} \ \rm{erg \ s^{-1}}$, which is concentrated within the UV band ($\sim 60$ eV). The X-ray ($0.3$ keV) flux of the SBO emission is about $0.002\mu$Jy, which can be easily submerged by the bright X-ray afterglow emission of the GRB of $\sim0.1\mu$Jy.
The Rayleigh-Jeans tail of the SBO emission produced the first peak in the UV-optical light curves of SN 2006aj, which can be much higher than or comparable to the luminosity of the main supernova peak. This result is very different from the situation of SLSNe, where the wind-driven SBO emission is always dimmer than the supernova emission \citep{2016ApJ...821...36K,2021ApJ...911..142L}. This difference can be understood by realizing that the remnant magnetars in long GRBs and SLSNe can distinguish each other by their different magnetic fields \citep{2017ApJ...840...12Y}. According to the obtained value of $L_{\rm sd,0}$ and $t_{\rm sd}$ for SN 2006aj, the magnetic field strength and initial spin period of its remnant magnetar can be derived to $B_{\rm p} =1.0\times10^{15}$ G and $P_{0}=1.0$ ms, which are both typical for long GRB magnetars as revealed from the plateau afterglow fittings \citep{2014ApJ...785...74L,2015MNRAS.454.3311M,2021MNRAS.508.2505Z}.

The mass of the SN ejecta is constrained to be $\sim 3.6M_{\odot}$, which is somewhat higher than the result ($\sim 2M_{\odot}$) obtained by \cite{2006Natur.442.1018M}. Meanwhile, as a result of the injection of the spin-down energy, the velocity of the supernova ejecta can finally approach to \citep{2016ApJ...819..120L}
\begin{equation}
v_{\rm ej}\sim \left({2L_{\rm sd,0}t_{\rm sd}\over M_{\rm ej}}\right)^{1/2}=23,000~\rm km~s^{-1},\label{vej}
\end{equation}
which is also roughly consistent with the velocity inferred by \cite{2006Natur.442.1018M} from the supernova spectra. Nevertheless, the difference is that \cite{2006Natur.442.1018M} considered only a small fraction of ejecta has such a high velocity, but our result indicates this high velocity could be an average value of the entire ejecta, which could correspond to an uncommon velocity profile for SN 2006aj \citep{2015ApJ...807..172N}. For the adopted opacity $\kappa=0.1\rm cm^{2}g^{-1}$, the obtained mass and velocity of the supernova ejecta determine a photon diffusion timescale as
\begin{equation}
t_{\rm diff}\sim \left({\kappa M_{\rm ej}\over 4\pi cv_{\rm ej}}\right)^{1/2}=10.5\rm ~day,
\end{equation}
which is much longer than the obtained spin-down timescale of the magnetar. Therefore, the injected energy is primarily converted into the kinetic energy of the ejecta rather than contributing to the supernova emission, as supposed in Eq. (\ref{vej}). This is the basic reason leading the GRBSN to be different from SLSNe \citep{2017ApJ...840...12Y}.

Nevertheless, Figure \ref{bol1} also shows that, although the spin-down luminosity had decreased significantly at the diffusion timescale, the contribution of the magnetar to the supernova emission can still be comparable to and even higher than the radioactive power for the adopted $M_{\rm Ni}=0.1M_{\odot}$. In the traditional radioactivity model for GRBSNe, a relatively high mass of $\rm ^{56}Ni\gtrsim 0.2 \  M_{\odot}$ is usually required to explain GRBSNe \citep{2018ApJ...862..130L}, since their luminosities are usually somewhat higher than normal core-collapse supernovae \citep{2016MNRAS.457..328L,2016MNRAS.458.2973P}. However, in this scenario, how such a huge amount of $\rm ^{56}Ni$ can be synthesized is an open question (see, \citealt{2009MNRAS.394.1317M,2015MNRAS.451..282S,2015ApJ...810..109N} and the references therein). In the magnetar engine model, in principle, the $\rm ^{56}Ni$ synthesization can be enhanced by the wind-driven shock \citep{2015MNRAS.451..282S}, which however usually requires a too high magnetic field to be consistent with the results inferred from the GRB afterglows. Therefore, as shown by our result, another possible explanation can be that the GRBSN emission is powered by substantial hybrid energy sources including the radioactivity and the magnetar engine, as also suggested by \cite{2016ApJ...831...41W,2017ApJ...837..128W} for SNe Ic-BL. In this case, we only need to invoke a relatively normal nickel mass.

Finally, two special parameter values are worth to be discussed. One is the initial value of the maximum radius of the ejecta which is found to be close to $R_{\rm max,0}\sim10^{14}$ cm, and another one is the gap $t_{\rm gap}$ between the time zeros of our calculation and the GRB trigger. The relatively large value of $R_{\rm max,0}$, which had also been realized in previous studies \citep{2008ApJ...683L.135C,2015ApJ...807..172N,2016MNRAS.460.1680I}, may indicate the progenitor star of SN 2006aj ejecta has a special extended structure. Alternatively, this may hint that it needs to spend a period of time to become a magnetar for the remnant NS (e.g., \citealt{2014ApJ...786L..13C}). So, during this period, the supernova ejecta can expand significantly. No matter whether this time gap is true or not, a time gap between the magnetar formation and the GRB trigger is definitely necessary for reproducing the rising phase of the SBO peak of the light curves. Specifically, the GRB trigger (i.e., the time zero of the data) is found to lag behind the magnetar formation (i.e., the time zero of our calculation) by a time of
\begin{equation}
t_{\rm gap}=t_{\rm GRB,0}-t_{\rm mag,0}=0.27\rm ~day.
\end{equation}
The possible existence of a time lag between the associated supernova and GRB had been previously noticed by some authors. For example, \cite{2003ApJ...588L..25F} expected this time lag to be at least larger than $\sim 30 \  \mathrm{s}$, \cite{2004AIPC..727..412N} suggested to be about $\pm 0.5 \ \mathrm{day}$, and \cite{1998Natur.395..672I} constrained it to be within $-0.7/+2 \ \mathrm{days}$ for GRB 980425/SN 1998bw. Our result demonstrates that the observations of the SBO rise can provide a practicable method to measure this time gap, which is crucial for testing the GRB mechanisms. However, in most cases, the UV-optical breakout can be easily overwhelmed by the GRB afterglow emission. For GRB 060218/SN 2006aj, its relatively faint afterglow and near distance make it possible to probe the detail of the UV-optical peak at very early time.

\section{Possible non-thermal PWN emission} \label{sec:PWNmodel}

\begin{figure*}
	\centering
	\includegraphics[width=0.49\textwidth]{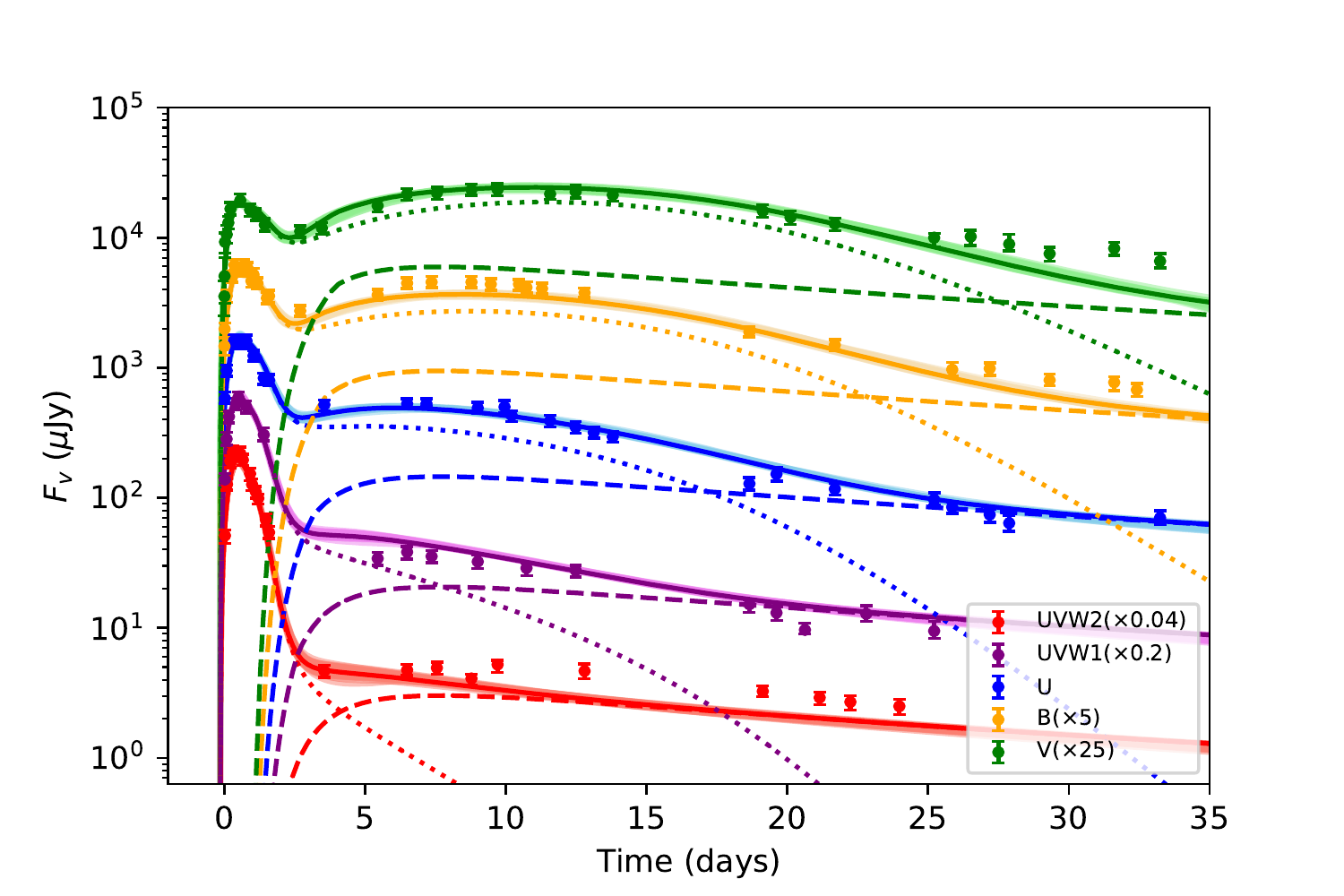} \includegraphics[width=0.49\textwidth]{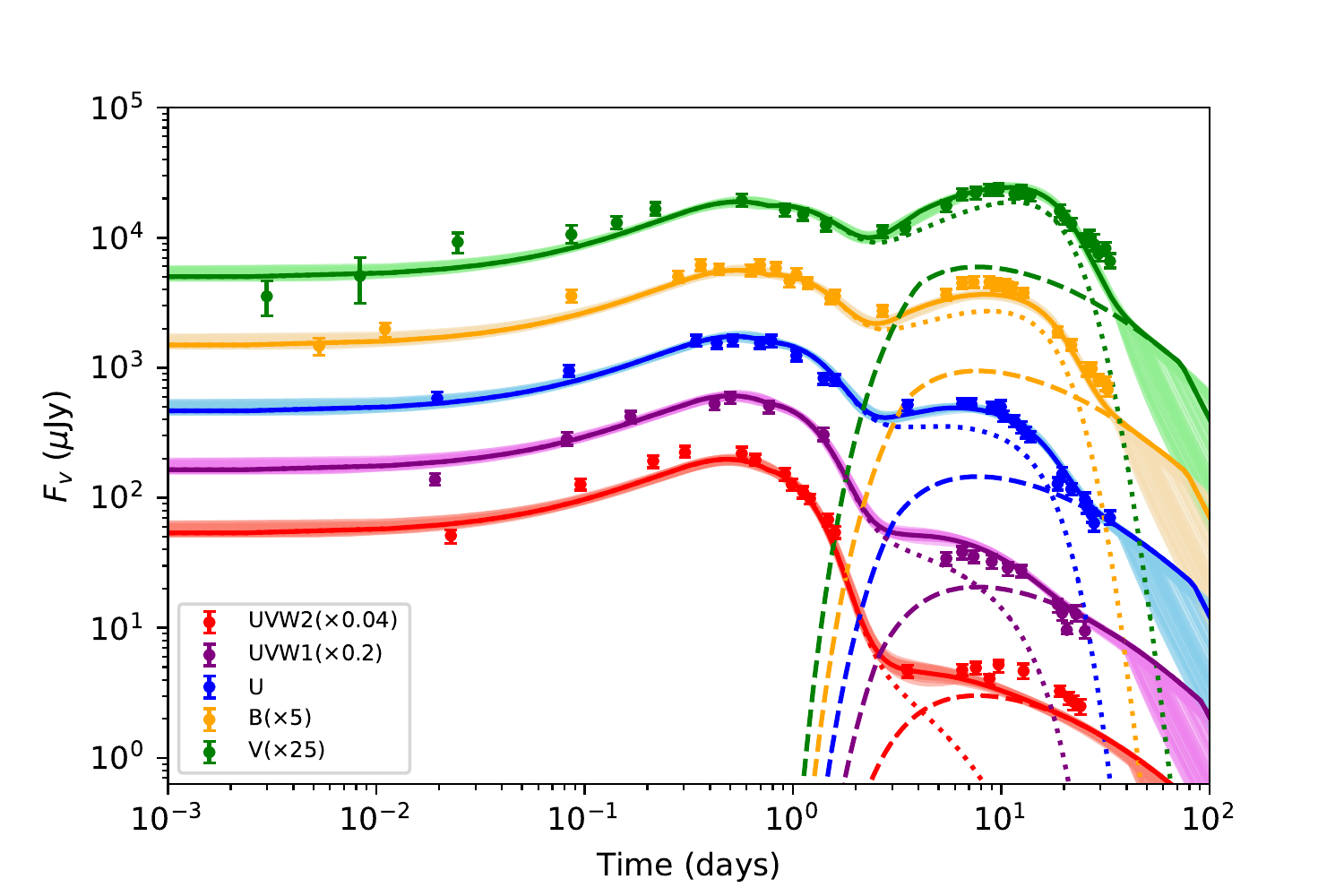}\\
	\caption{The same as Figure \ref{LC1} but with an extra non-thermal PWN emission component shown in dashed lines, the thermal component is shown in dotted lines.}
	\label{mul2}
\end{figure*}

Although the fittings presented in the previous section account for the basic features of SN 2006aj, the observational data exceed the theoretical curves significantly for relatively high frequencies, especially in the UV bands. Such a discrepancy between the model and data suggests that the spectra of SN 2006aj deviate from an ideal black body in the UV bands, which becomes more serious at later time. This UV excess could be a result of frequency-dependent radiative transfer and could also indicate an extra non-thermal emission component. In the magnetar engine model, such a non-thermal emission is most likely to originate from the relativistic wind of the magnetar behind the supernova ejecta. When the wind collides with the supernova ejecta, a termination shock can be driven and propagate into the wind. As usual, we termed the shocked wind region as a pulsar wind nebula (PWN), which can in principle contribute to a significant non-thermal emission after the supernova ejecta gradually becomes transparent \citep{2013MNRAS.432.3228K,2019ApJ...877L..21Y,2021A&A...654A.124W}. Then, in this section, we try to refit the observational data of SN 2006aj by invoking such a non-thermal PWN emission.

We calculate the PWN emission by using the model presented in \cite{2019ApJ...877L..21Y}, where the most crucial step is to determine the comoving density of the internal energy of the PWN as
\begin{equation}
	e'_{\rm pwn}=4\Gamma_{\rm ts}^{\prime \ 2} n^{\prime}_{\rm w}m_{e}c^2,
	\label{epwn}
\end{equation}
where $\Gamma^{\prime}_{\rm ts}=\frac{1}{2}\Gamma_{\rm w}$ is the Lorentz factor of the termination shock relative to the unshocked wind of a Lorentz factor $\Gamma_{\rm w}$ and
\begin{equation}
	n^{\prime}_{\rm w}= \frac{L_{\rm sd}}{4\pi R_{\rm ts}^2 \Gamma_{\rm w}^2 m_{e} c^3}
\end{equation}
is the comoving number density of the unshocked wind which is considered to purely consist of electron and positron pairs. Here, the luminosity of the relativistic wind is simply given by the spin-down luminosity of the magnetar, and $R_{\rm ts}$ is the radius of the termination shock. In our calculation, the value of $e'_{\rm pwn}$ can be directly obtained from the mechanical equilibrium between the PWN and the supernova ejecta, i.e.,
\begin{equation}
	e'_{\rm pwn}=e'_{\rm ej}={3\tilde{U}\over 4\pi R_{\rm sh}^3}\label{epwn2}.
\end{equation}
Then, by introducing equipartition factors $\epsilon_{\rm B}$ and  $\epsilon_{\rm e}$ for the magnetic field and electrons in the PWN, and assuming the electron energy distribution with a power-law index $p$.  Thus we can calculate the synchrotron emission of these electrons (see \citealt{1998ApJ...497L..17S,2019ApJ...877L..21Y} for details).

For a PWN synchrotron luminosity of $L_{\nu,\rm pwn}$, the luminosity of the leaking non-thermal component appearing in the supernova emission can be given by
\begin{equation}
	L_{\nu,\rm nth}=L_{\nu,\rm pwn} {\rm e}^{-\tau_{\nu}}.\label{Lnth}
\end{equation}
Simultaneously, the power absorbed by the supernova ejecta can be written as
\begin{equation}
	L_{\rm sd,inj}=\int L_{\nu,\rm{pwn}}(1-e^{-\tau_{\nu}})d\nu, \label{Linj}
\end{equation}
which would be used in the previous section for calculating the dynamical evolution and emission of the supernova ejecta. Here, the optical depth includes both contributions from the shocked and unshocked supernova ejecta, which reads
\begin{eqnarray}
\tau_{\nu}&=&\tau_{\rm sh,\nu}+\tau_{\rm un,\nu}\nonumber\\
&=& \frac{3 \kappa_{\nu} M_{\rm ej}}{4\pi R_{\rm sh}^{2} }+ \int_{R_{\rm sh}}^{R_{\max}} \kappa_\nu \rho_{\rm ej}
dr,
\end{eqnarray}
where the frequency dependence of the opacity is taken into account because the non-thermal PWN emission can appear in a wide range of electromagnetic bands. By fitting the numerical result presented in \cite{2013MNRAS.432.3228K}, \cite{2019ApJ...877L..21Y} suggested
\begin{eqnarray}
\kappa_{\nu}&=&
		C_{1}\left(\frac{h \nu}{1 \  \rm{keV}}\right)^{-3}+ C_{2}\left[\left(\frac{h \nu}{50 \  \rm{keV}}\right)^{-0.36}\right.\nonumber\\
&&\left.+\left(\frac{h \nu}{50 \ \rm{keV}}\right)^{0.65}\right]^{-1} + C_{3}
\end{eqnarray}
for $h \nu \gtrsim 0.1 \  \rm{keV}$, where the coefficients are $C_{1}=2.0 \ \rm{cm^2g^{-1}}$, $C_{2}=0.7 \ \rm{cm^2g^{-1}}$, and $C_{3}=0.012 \ \rm{cm^2g^{-1}}$.

\begin{figure*}[htbp]
	\centering
	\includegraphics[width=1\textwidth]{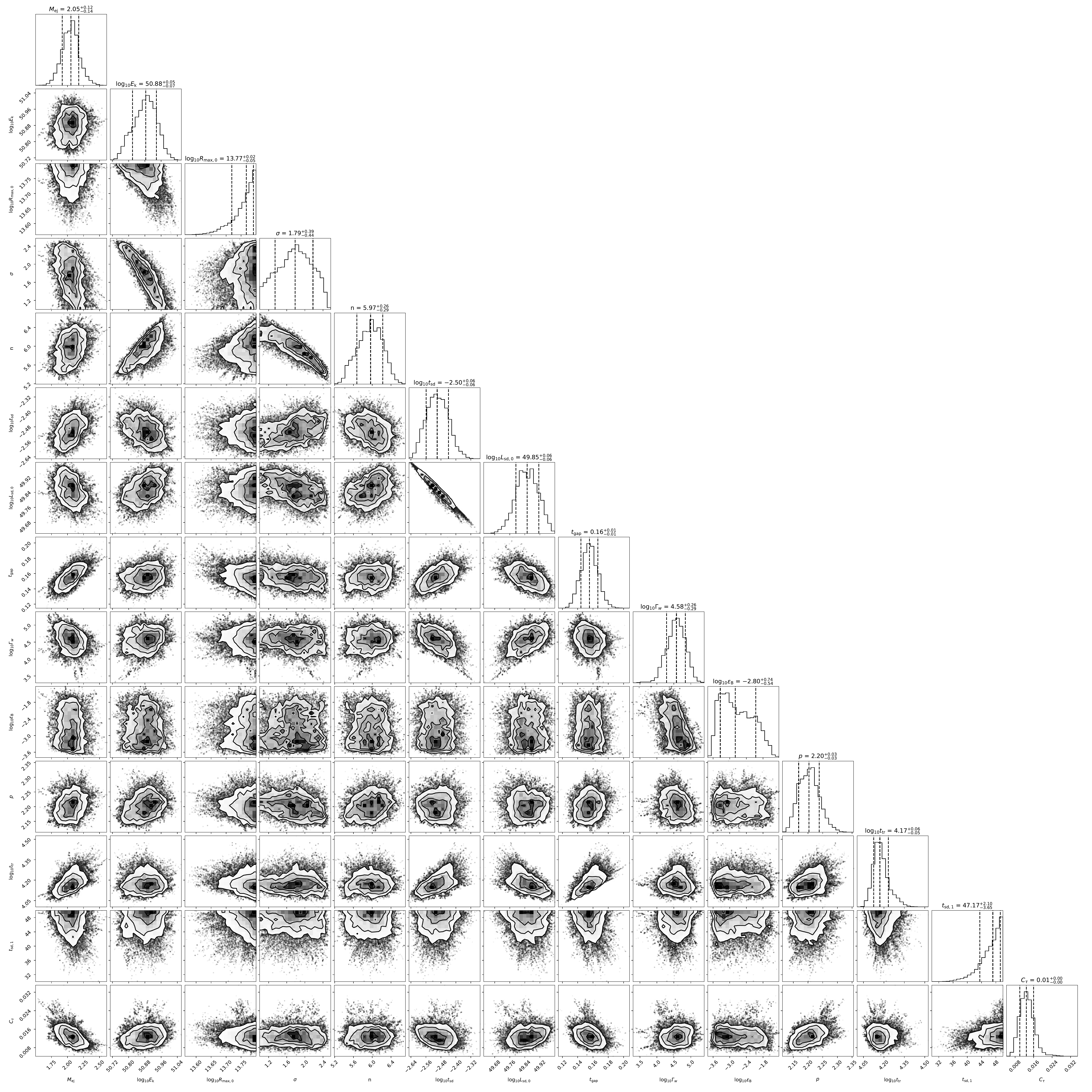} \\
	\caption{Posterior distributions of parameters for the fittings of SN 2006aj with the model in Section \ref{sec:PWNmodel}.}
	\label{MCMC2}
\end{figure*}

\begin{table}[htbp]
	\centering
	\setlength{\tabcolsep}{2mm}{}
	\renewcommand\arraystretch{1.4}
	\begin{tabular}{lllll}
		\hline \hline Parameter & Value & Best-fit & Range \\
		\hline
		$M_{\rm ej} \ (M_{\odot})$ &  $2.05_{-0.14}^{+0.12} $ & 1.92  & (1,10)\\
		$\log_{10}E_{\rm k} \ (\rm{erg})$ & $50.88_{-0.07}^{+0.05}$ & 50.89 & (50,52) \\
		$\log_{10}R_{\rm{max,0}} \  (\rm{cm})$&   $13.77_{-0.05}^{+0.02}$ & 13.78 & (11,13.8) \\
		$\sigma$ & $1.79_{-0.44}^{+0.39}$ & 1.78 & (1,3)\\
		$n$ & $5.97_{-0.29}^{+0.26}$ & 6.04 & (5,11)\\
		$\log_{10}t_{\rm sd} \  (\rm{day})$ & $-2.50_{-0.06}^{+0.06}$ & -2.56 & (-4,0) \\
		$\log_{10}L_{\rm sd,0} \ (\rm{erg/s})$ & $49.85_{-0.06}^{+0.06}$ & 49.92 & (47,50)\\
		$t_{\rm gap} \ (\rm{day})$ &  $0.16_{-0.01}^{+0.01}$ & 0.14 & (0,2)\\
		$\log_{10}\Gamma_{\rm w}$ & $4.58_{-0.29}^{+0.26}$ & 4.69 & (2,7)\\
		$\log_{10}\epsilon_{\rm B}$ & $-2.80_{-0.54}^{+0.74}$ & -2.46 & (-6,-1)\\
		$p$ & $2.20_{-0.03}^{+0.03}$ & 2.16 & (2,3)\\
		$\log_{10}{t_{\rm tr}} \ (\rm{s})$ & $4.17_{-0.05}^{+0.06}$ & 4.09 & (4,6)\\
		$t_{\rm sd,1} \ (\rm{day})$ & $47.17_{-3.65}^{+2.10}$ & 49.10 & (1,50)\\
		$C_{\rm \tau}$ & $0.01_{-0.00}^{+0.00}$ & 0.01 & (0,1)\\
		$M_{\rm Ni} \ (M_{\odot})$ & 0.1 (fixed) \\
		
		\hline
		$E_{\rm sd} \ (\rm{erg})$ &  & $1.979 \times 10^{52}$ & \\
		$p_{\rm sd,0} \ (\rm{ms})$ &  & 1.022 & \\
		$B_{\rm p} \  (10^{14}\rm{G})$ &  & 15.058 & \\
		\hline
		
	\end{tabular}\\
	
	\caption{Parameters estimated from modeling the double-peaked light curve of SN 2006aj with the model in Section \ref{sec:PWNmodel} by MCMC-sampling. Where $t_{\rm sd,1}$ represents the spin-down timescale after the magnetic dipole radiation suppression.} \label{tab2}
\end{table}

\begin{figure}[htbp]
	\centering
	\includegraphics[width=1\textwidth]{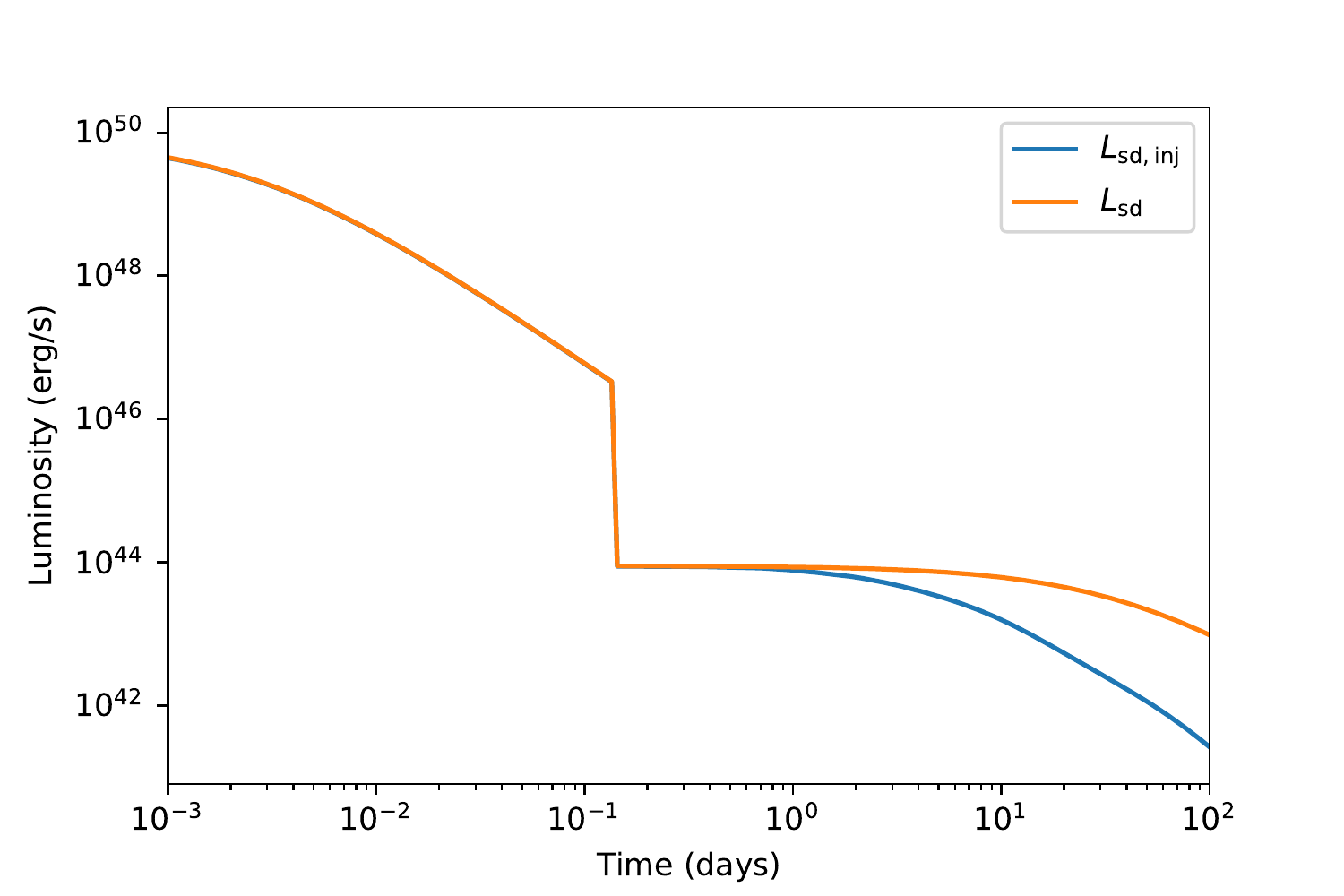} \\
	\caption{The temporal evolution of the spin-down luminosity of the mangetar, which is required by the consistency of the non-thermal PWN emission with the observational data of SN 2006aj. The zero-time in this figure is $t_{\rm mag,0}$. }\label{Lsd2}
\end{figure}

Combining Eqs. (\ref{Lsbo}), (\ref{Lsn}) and (\ref{Lnth}), we refit the light curves of SN 2006aj and present the result in Figure \ref{mul2}. The corresponding parameter values are shown in Table \ref{tab2} and Figure \ref{MCMC2}. As shown, a plausible fit can be obtained if two ad hoc assumptions are made, where the excesses in the UV bands are successfully accounted for and the supernova's rising phase is better fitted. On the one hand, it is assumed the magnetic dipole radiation of the magnetar is suddenly suppressed at a time of $t_{\rm tr}$ due to an unknown reason, the luminosity evolution of which is shown in Figure \ref{Lsd2}. Although it seems unusual, such a sudden change has been previously suggested to explain the overall evolution of short GRB 170817A and its associated kilonova AT 2017gfo \citep{2018ApJ...861..114Y}. In principle, the suppression of the magnetic dipole radiation could be caused by a fallback accretion that buries the surface dipolar magnetic field of the magnetar \citep{2021ApJ...907...87L}. On the other hand, an artificial suppression on the optical depth $\tau_{\nu,\mathrm{sup}}=C_{\tau} \tau_{\nu}$ should be invoked in Eq. (\ref{Lnth}) and (\ref{Linj}), so that the PWN emission can leak not too late. This hints the first leaked PWN emission could come from a radius much larger than $R_{\rm ts}$ and very close to the surface of the supernova ejecta. Such a situation could appear if the supernova ejecta is highly anisotropic, which seems not strange in the extreme stellar explosive phenomena such as GRBs. In addition, as illustrated in Figure \ref{ske1}, the early leakage of the PWN emission could also be due to turbulence of the interface between the PWN region and the supernova ejecta, which is subject to the Rayleigh-Taylor instabilities \citep{2004A&A...423..253B,2005ApJ...619..839C,2017hsn..book.2159S}. In this case, a remarkable fraction of the PWN material (the `mushroom finger') can penetrate into the supernova ejecta deeply.

\section{Conclusion and discussion} \label{sec:summary}
In this paper, we revisit the famous double-peaked GRBSN, SN 2006aj, to investigate its consistency with a millisecond magnetar engine. The effects of the magnetar on the supernova emission can be in principle shown in the following three aspects. (i) The powerful magnetar wind can drive a forward shock crossing the supernova ejecta, which can accumulate a remarkable amount of internal energy in the region immediately behind the shock front. As the ejecta material outside the shock becomes transparent, the shock breaks out and produces a transient emission. The Rayleigh-Jeans tail of this SBO emission can well account for the first peaks in the UV-optical light curves of SN 2006aj, which is the most remarkable feature of this supernova. (ii) Besides the traditional radioactive power, the primary peak of the supernova emission can also be powered by the energy injection from the magnetar wind, although the spin-down luminosity had been reduced significantly at $t_{\rm diff}$ which is much longer than $t_{\rm sd}$. The extra power leads the GRBSN to be somewhat more luminous than normal core-collapse supernovae, even though not too many $^{56}$Ni is synthesized. (iii) As a result of the interaction between the magnetar wind and the supernova ejecta, the energy carried by the wind would be first converted into PWN emission, which can be effectively absorbed by the supernova ejecta at early times. Nevertheless, as the expansion of the supernova ejecta, the PWN emission can finally leak from the ejecta to be observed, which then contributes a non-thermal component to the late supernova emission. This component could provide an explanation for the UV excess in the SN 2006aj emission.

The good agreement between the theoretical expectation and observational features demonstrates that the central engine of GRB 060218/SN 2006aj is very likely to be a magnetar, with a magnetic field of $B_{\rm p} \sim 10^{15}$ G. This magnetic field strength is a typical value for usual GRB magnetars as inferred from afterglow emission, which indicates the self-consistence of the magnetar engine model. Meanwhile, this magnetic field strength is also the key to understand the difference between the GRBSNe and the SLSNe. In addition, two ad hoc treatments in our calculations need to be emphasized. On the one hand, it is strongly suggested that time gaps can exist between the supernova explosion, the magnetar formation, and the GRB trigger, which can offer crucial information for exploring the GRB mechanisms. In the future, besides to find more double-peaked GRBSNe like SN 2006aj, multi-messenger observations including gravitational wave and neutrino emission would make us be able to more directly measure the time lags \citep{2002ApJ...565..430F,2003ApJ...584..971B,2004BaltA..13..317M,2008MNRAS.385.1461Y}.
On the other hand, the geometry of the GRBSN ejecta could be very different from a simple isotropic structure and the interface between the magnetar wind and the ejecta could also seriously deviate from a simple spherical surface (e.g., due to the Rayleigh-Taylor instabilities). Only invoking these complexities, the non-thermal component appearing in the SN 2006aj emission at the late time can be explained by the leakage of the PWN emission.

\begin{acknowledgments}
This work is supported by the National Key R\&D Program of China (2021YFA0718500), the China Manned Spaced Project (CMS-CSST-2021-A12), the National SKA program of China (2020SKA0120300), and the National Natural Science Foundation of China (Grant No. 11822302 and 11833003).
\end{acknowledgments}

\bibliography{grbsne}{}
\bibliographystyle{aasjournal}

\end{document}